\documentclass[12pt]{iopart}
\usepackage{iopams}
\usepackage{enumitem}
\usepackage{graphicx}
\usepackage{braket}
\usepackage{float}
\usepackage{cite}
\usepackage{pdflscape}
\usepackage{afterpage}
\usepackage{soul}
\newcommand{\nx}{\noindent}
\newcommand{\beq}{\begin{equation}}
\newcommand{\eeq}{\end{equation}}
\newcommand{\beqa}{\begin{eqnarray}}
\newcommand{\eeqa}{\end{eqnarray}}
\newcommand{\beqn}{\begin{eqnarray*}}
\newcommand{\eeqn}{\end{eqnarray*}}
\newcommand{\aver}[1]{\big\langle #1 \big\rangle}
\newcommand{\av}[1]{\langle #1\rangle}

\newcommand{\lamscrs}[2]{\Lambda_{#1}^{#2}}
\newcommand{\rhoa}{\rho_{\scriptsize \textsc{a}}}
\newcommand{\p}[2]{p_{#1}^{(#2)}(x)}
\graphicspath{{./figures/}}

\begin{document}

\title{Exact eigenvalue order statistics for the reduced density 
matrix of a bipartite system}

\author{B. Sharmila$^{1,2}$, V. Balakrishnan$^{1}$ and  S. Lakshmibala$^{1}$}

\address{$^{1}$ Department of Physics, Indian Institute of Technology Madras, Chennai 600036, India.\\
$^{2}$ Present address: Department of Physics, University of Warwick, Coventry CV4 7AL, UK.}
\vspace{10pt}
\begin{indented}
\item[]\today
\end{indented}

\begin{abstract}
We  consider the reduced density matrix 
$\rhoa^{(m)}$ of a bipartite 
system $AB$ of dimensionality 
$mn$ in  a Gaussian ensemble of 
random, complex  pure states of the 
composite system.
For a given dimensionality $m$ of the subsystem $A$,
 the eigenvalues  
$\lambda_{1}^{(m)},\ldots, 
\lambda_{m}^{(m)}$  of $\rhoa^{(m)}$ are correlated 
random variables because their sum equals 
unity. The following quantities are 
known, among others: The joint probability density function (PDF) 
of the eigenvalues 
$\lambda_{1}^{(m)},\ldots, 
\lambda_{m}^{(m)}$  of $\rhoa^{(m)}$, 
the PDFs of  
the smallest eigenvalue 
$\lambda_{\rm min}^{(m)}$ 
and the largest eigenvalue $\lambda_{\rm max}^{(m)}$, 
and the family of average values
$\aver{\mathrm{Tr}\big(\rhoa^{(m)}\big)^{q}}$ 
parametrised by $q$. 
Using values of $m$ running from 
$2$ to $6$ for definiteness,   
we show that these inputs suffice to 
identify and characterise the eigenvalue  
order statistics,  i.e., 
to obtain 
explicit analytic expressions for the PDFs of each  of the $m$ eigenvalues arranged in ascending order 
from the smallest to the largest one. 
When $m = n$ (respectively, 
$m < n$)   these PDFs are polynomials of order 
$m^{2}-2$ (respectively, 
$mn - 2$) 
 with support in specific sub-intervals of the unit interval, demarcated by appropriate 
unit step functions.  
Our exact results are 
fully corroborated by numerically generated histograms 
of the ordered set of eigenvalues corresponding to ensembles of over $10^{5}$ random 
complex pure states of the bipartite system.  
 Finally, we present the general solution 
for arbitrary values 
of the subsystem dimensions 
 $m$ and $n$, namely,  
 formal exact expressions for 
 the PDFs of every ordered eigenvalue.
\end{abstract}
\vspace{2pc}
\noindent{\it Keywords}: Bipartite system, random complex pure states, Gaussian ensemble, 
reduced density matrix,  eigenvalue order statistics.

\maketitle
%
%

\section{Introduction \label{sec:intro}}

A basic motif 
in the study of entanglement involves a 
bipartite system $AB$ 
comprising subsystems $A$ and $B$ with  Hilbert space dimensions $m$ and $n (\geqslant m)$, respectively. Given a specified Gaussian  ensemble of random pure states  of the composite system, 
the task is to deduce the statistical properties 
of the reduced density matrix $\rhoa^{(m)}$. 
A fairly  extensive literature exists in this regard
(see Refs. \cite{lubkin}--\cite{bianchi} and references therein).
The  properties investigated include the 
joint probability density function (PDF) 
 of the  set of eigenvalues $\{\lambda_{k}^{(m)}\}$ ($1\leqslant k \leqslant m$) 
 of $\rhoa^{(m)}$ \cite{lubkin,lloyd}, the leading large-$m$ behaviour of 
the average values of the set of eigenvalues~\cite{znidaric},
the PDFs of the smallest 
 and largest eigenvalues 
$\lambda_{\rm min}^{(m)}$ \cite{edelman,znidaric,majumdar,chen}  and
$\lambda_{\rm max}^{(m)}$ \cite{vivo},   
their mean values and higher moments 
$\aver{(\lambda_{\rm min}^{(m)})^{q}}$ 
and $\aver{(\lambda_{\rm max}^{(m)})^{q}}$, 
the average 
$\av{\mathrm{Tr}\,(\rhoa^{(m)})^{q}}$ where 
$q$ is any positive integer \cite{majumdar,vivo,bianchi}, 
and the associated entropies that quantify the 
extent of entanglement of $A$ and $B$ 
such as the average subsystem von Neumann  entropy 
$-\aver{\mathrm{Tr}(\rhoa^{(m)} \ln \rhoa^{(m)})}$ \cite{page}--\cite{sen}  and the subsystem linear entropy 
$1 - \av{\mathrm{Tr}\,(\rhoa^{(m)})^{2}}$ \cite{lubkin}.
 The main 
technical complication in these studies arises 
from the fact that the 
eigenvalues are   correlated random variables.

The studies listed  in the foregoing deal,  by and large,  with the {\em set} of eigenvalues of 
$\rhoa^{(m)}$ rather than 
individual ones, with the exception of the extreme values
$\lambda_{\rm min}^{(m)}$ 
and $\lambda_{\rm  max}^{(m)}$ 
as already stated.
The natural extension 
of extreme value statistics  
is the {\em order statistics}  of the  
eigenvalues, the task being to deduce  the 
PDFs of the individual eigenvalues 
identified by their positions in an 
{\em ordered}  
sequence. This is the purpose of the 
present paper.  
 The complications involved in 
the statistics of extreme values in the case 
of correlated random variables persist, of course, 
for order statistics as well. In order to avoid any confusion,  we   shall denote the eigenvalues 
$\lambda_{1}^{(m)},  \lambda_{2}^{(m)}, 
\ldots, \lambda_{m}^{(m)}$ arranged 
in ascending order by the sequence 
\beq
\lamscrs{1}{(m)}, 
\lamscrs{2}{(m)}, 
\ldots, \lamscrs{m}{(m)}.
\label{evsinorder}
\eeq
Thus $\lamscrs{1}{(m)}\equiv \lambda_{\rm min}^{(m)}$ and 
$\lamscrs{m}{(m)}\equiv \lambda_{\rm max}^{(m)}$. We seek the normalized PDF 
$\p{k}{m}$  (where $x \in [0, 1]$) 
corresponding to each 
$k^{\rm th}$-order statistic $\lamscrs{k}{(m)}$,  where $1 \leqslant k \leqslant m$. 
The investigations cited   
in the opening   paragraph above 
rely  on the following 
basic result \cite{lloyd}. 
Let  $\lambda_{1}^{(m)}, \ldots, \lambda_{m}^{(m)}$ be the eigenvalues of $\rhoa^{(m)}$ 
 (listed in no particular order). Their joint PDF is 
then  given by 
\beqa
\nonumber P(\lambda_{1}^{(m)},\lambda_{2}^{(m)},\dots,\lambda_{m}^{(m)})=C_{m, n}^{(\beta)} \,\delta &\Big( \sum_{i=1}^{m} \lambda_{i}^{(m)} - 1 \Big) \prod_{i=1}^{m} \big(\lambda_{i}^{(m)}\big)^{\alpha}\\
& \times \prod_{j<k} 
|\lambda_{j}^{(m)}-\lambda_{k}^{(m)}|^{\beta}.
\label{jointprobability}
\eeqa
Here   the Dyson index 
$\beta = 2$ (respectively, $1$)  
 for  a Gaussian ensemble  of complex 
 (respectively,  real)  pure states, 
 \beq
 \alpha=(\beta/2)(n-m+1)-1,
 \label{alphadefn}
 \eeq 
 and
the normalizaton constant is 
\beq
C_{m, n}^{(\beta)}=\frac{\Gamma(m n \beta/2) \left[\Gamma(1+(\beta/2))\right]^{m}}{\prod\limits_{j=0}^{m-1} \Gamma((n-j)\beta/2)\Gamma(1+(m-j)\beta/2)}.
\label{normalizationC}
\eeq
As is well known in random matrix theory, 
the eigenvalues $\{\lambda_{k}^{(m)}\}$ form  a set 
of correlated random variables both because of the  
bunching effect  arising from the requirement that their sum be equal to unity, as well as 
the level repulsion implied by the presence of the 
factor $|\lambda_{j}^{(m)}-\lambda_{k}^{(m)}|^{\beta}$ 
in the joint PDF of Eq. (\ref{jointprobability}). 
In principle, the PDF $\p{k}{m}$ of the $k^{\rm th}$ eigenvalue in the ordered eigenvalue sequence
can be found by multiplying the joint 
PDF in Eq. (\ref{jointprobability}) by the product 
of step functions 
\beq
\prod_{j=1}^{k-2}
\Theta (\lambda_{j+1}- \lambda_{j})\,
\Theta( x - \lambda_{k-1})
\Theta(\lambda_{k+1}- x)\,
\prod_{l=k+1}^{m-1}
\Theta(\lambda_{l+1}- \lambda_{l}),
\label{thetaproduct}
\eeq
integrating over $\lambda_{1},  \ldots,
\lambda_{k-1}, \lambda_{k+1}, 
\ldots, \lambda_{m}$, and normalizing the 
resulting function of $x$ to unity. 
This approach, however, presents 
formidable technical problems, and 
is not feasible. We shall see that there is 
an alternative,  simpler procedure to arrive at 
the result sought.   \\

It is evident 
from Eqs. (\ref{jointprobability}) and 
(\ref{normalizationC}) 
that some simplification occurs when $\beta = 2$ 
(complex pure states), and we shall consider this case. 
While we shall finally present exact expressions for the 
PDFs of the ordered eigenvalues for arbitrary  
subsystem dimensions $m$ and $n$, those expressions 
are somewhat complicated. The manner in which the 
structure of the solutions arises   is best elucidated 
 by explicit illustration for small values of $m$.  
 Accordingly, we shall start with $m = 2$ and 
 increase it step by step up to $m = 6$, to 
 demonstrate  how the 
 PDFs build up.  In the interests of clarity, we shall 
 further set $n = m$ for the most part in this 
 demonstration, in order to take advantage of the 
 simplification that ensues from the fact that 
 $\alpha = 0$ in this case. \\ 
 
The plan of the paper is as follows. 
In Section \ref{sec:preliminaries}, 
we write down the ranges of  
the random variables  $\lbrace \lamscrs{k}{(m)} \rbrace$, followed by some properties of Mellin 
transforms that will be used in the sequel. We also 
list three specific, already  known results 
that will be used to deduce the PDFs 
of the ordered eigenvalues. Next, in Section \ref{sec:mn2}
we state (for the sake of completeness)   
the existing results in the trivial case $m=n=2$. 
 We then discuss in Section \ref{sec:mn3}
 the  case  $m=n=3$, which is again 
 special because   
   there is just one eigenvalue in between   the smallest and largest eigenvalues.  In Sections \ref{sec:mn4} and \ref{sec:others}, we illustrate our procedure to obtain the analytical expressions for PDFs corresponding to $m=4$, $5$, and $6$ 
   when  $m=n$. We also show that our procedure works for the case $m \neq n$ by considering the case $m=4$, $n=5$. Finally, in Section \ref{sec:generalmn} we 
   present the solution for the PDFs 
   $\{\p{k}{m}\}, \,1\leqslant k \leqslant m$, for arbitrary 
   values of the subsystem dimensions $m$ and 
   $n \geqslant m$.   
   We conclude with brief remarks on the behaviour of the 
 PDFs  at the end points of their 
 domains.

\section{Preliminaries}
\label{sec:preliminaries}
Integrating 
$P(\lambda_{1}^{(m)}, \ldots, \lambda_{m}^{(m)})$ over all but any one of the set $\{\lambda_{k}^{(m)}\}$ is a technically complicated  task. It  yields a so-called 
`single-particle' PDF for 
a single eigenvalue
\cite{adachi}. But this  
procedure automatically averages over the location of the eigenvalue in the  ordered set
of eigenvalues. 
 Thus, it leads, for instance,  to the expected result that the average 
value $\aver{\lambda_{k}^{(m)}}$ is just $1/m$.
It is evident  that this single-particle PDF is 
quite different from the individual PDFs corresponding to the ordered eigenvalues.  \\

We start with some 
 general properties of the ordered set 
of eigenvalues 
$\lbrace \lamscrs{k}{(m)} \rbrace$ 
that will be needed to identify and 
isolate  the 
corresponding set of PDFs of the individual 
members of the set.
Since ${\rm Tr}\, \rhoa^{(m)} = 1$, the ordered  set of non-negative  numbers 
$\{\lamscrs{k}{(m)}\}$ satisfies the relation
$\sum_{k=1}^{m}\lamscrs{k}{(m)} 
= 1$.
It follows at once that 
$\lamscrs{1}{(m)}$ 
cannot exceed $1/m$, while 
$\lamscrs{2}{(m)}$ cannot exceed $1/(m-1)$, 
and so on. That is, 
\beq
0 \leqslant \lamscrs{k}{(m)} 
\leqslant  1/(m+1-k), \;\; k = 1,2,\ldots, m-1.
\label{evrange1}
\eeq
In particular,  $\lamscrs{m-1}{(m)}\leqslant 1/2$. 
Moreover, it is evident that  the largest eigenvalue cannot be {\em  smaller}  than $1/m$, so that its range 
is given by 
\beq
1/m \leqslant 
\lamscrs{m}{(m)} 
\leqslant 1.
\label{evrange2}
\eeq
The ranges in Eqs. (\ref{evrange1}) 
and (\ref{evrange2}) therefore specify the 
support $I_{k}^{(m)}$ of the normalized 
 PDF  $\p{k}{m}$ of each $\lamscrs{k}{(m)}$,  
 for  $1 \leqslant k \leqslant m$. \\
 
 As we shall also be concerned with 
 the higher moments 
 of the eigenvalues $\lamscrs{k}{(m)}$, it is helpful 
 to recall very briefly some properties of Mellin transforms.  
The Mellin transform 
$\widetilde{f}(q)$ 
of a function 
$f(x) \, (x \in [0,1])$ and its inverse 
are defined as
\beq
\widetilde{f}(q) 
= \int_{0}^{1} x^{q} f(x) dx, \;\;
f(x) = \frac{1}{2\pi i}
\int_{C}x^{-q-1}\widetilde{f}(q)dq,
\label{mellin}
\eeq
where the  
Bromwich contour $C$ runs
from $c-i\infty$ to $c+i\infty$ 
to the right of all the singularities of 
$\widetilde{f}(q)$. 
The following easily established results will be used in the sequel:  \\

\nx
(a)\, If $\widetilde{f}(q)$ is a rational function of 
$q$ with simple poles at the negative integers $q = -1, \ldots, -\nu$, then $f(x)$ is a polynomial in $x$ of 
order $\nu -1$, multiplied by the unit step function
$\Theta(1-x)$.\\

\nx
(b) Let $r$ be a positive integer $>1$. 
If $\widetilde{f}(q)$ is 
$r^{-q}$  times a rational function of 
$q$ with simple poles at the negative integers $q = -1, \ldots, -\nu$, then $f(x)$ is a polynomial in 
$x$ of order $\nu -1$, multiplied by the unit step function
$\Theta(1-rx)$.\\  

\nx
In the present context, 
we note that  
the  PDF $\p{k}{m}$ and  
the $q^{\rm th}$ moment of 
$\lamscrs{k}{(m)}$, namely, 
\beq
\aver{(\lamscrs{k}{(m)})^{q}}
\equiv \int_{I_{k}^{(m)}} x^{q} 
\p{k}{m} dx, 
\label{qthmoment}
\eeq
comprise a Mellin transform pair.  \\

We turn  
now to  the three  known results 
that we  require  to deduce the PDFs 
of the ordered eigenvalues, setting 
$\beta = 2$ and $m = n$. \\

\nx
(i)  The PDF $\p{1}{m}$ of the smallest eigenvalue 
$\lamscrs{1}{(m)}$ is given by \cite{majumdar}
\beq
\p{1}{m}=m(m^{2}-1) (1-m x)^{m^{2}-2} \Theta (1- m x),
\label{pdflambdamin}
\eeq
where $\Theta$ denotes the unit step function. 
The corresponding moments of the smallest eigenvalue are then given by  
\beq
\aver{(\lamscrs{1}{(m)})^{q}}= 
\frac{\Gamma(q+1) \Gamma(m^{2})}
{m^{q} \,\Gamma (m^{2}+q)}\,.
\label{momentslambdamin}
\eeq 

\nx
(ii)   The average 
$\aver{\mathrm{Tr} (\rhoa^{(m)})^{q}}$
 is given by \cite{bianchi}  
\beq
\aver{\mathrm{Tr} (\rhoa^{(m)})^{q}}=
\frac{\Gamma(m^{2})}{\Gamma(m^{2}+q)} \sum\limits_{i,p=0}^{m-1} \frac{\Gamma(p+q+1) [\Gamma(q+1)]^{2}}
{\left[\Gamma(1+i-p)\Gamma(1+q+p-i)\right]^{2} p!}.
\label{tracerhoq}
\eeq
We  note that this average  is the Mellin transform of the  sum of PDFs, 
$\sum_{k=1}^{m}
\p{k}{m}$. Inverting the transform will therefore 
yield that sum. \\

\nx
(iii) The third and most crucial ingredient is the 
 PDF $\p{m}{m}$ 
of the largest 
eigenvalue $\lamscrs{m}{(m)}$, 
for which an implicit formula  has been 
derived in Ref. \cite{vivo}. As pointed out therein, 
the determination of $\p{m}{m}$ involves several technical 
complications that are not present in the 
determination of $\p{1}{m}$. 
The procedure \cite{vivo}  leading to the result sought may be summarized 
in the following sequence of steps.
  First, one defines the  
set of $m^{2}$ functions 
\beq
 \Psi_{jl}(s)= \int \limits_{0}^{1} 
 e^{-s u} u^{j+l} du \;\; (j, l = 0, 1, \ldots, m-1)
\label{psijl}
\eeq
and evaluates the $(m \times m)$ determinant 
\beq
 \widetilde{\mathcal{P}}(s)
 =\mathrm{det}\left[ \Psi_{j l}(s) \right]_{j,l=0,1,\dots,m-1}.
\label{Ptilde}
\eeq
The inverse  Laplace 
transform $\mathcal{P}(t)$ 
of $\widetilde{\mathcal{P}}(s)$ 
is then obtained, and  
$t$ is set equal to  $1/x$ 
in the result. 
Then, the quantity
\beq
Q_{m}(x) = 
(m^{2} -1)! \prod_{j=0}^{m-1}
\frac{(j+1)! (j!)^{2}}
{(m-1-j)! (m-j)! (m+j)!} x^{m^{2}-1}
\mathcal{P}(1/x)
\label{cdfoflambdamax}
\eeq
is   the cumulative distribution function  
of $\lamscrs{m}{(m)}$, 
from which its PDF follows
according to 
\beq
\p{m}{m}= dQ_{m}(x)/dx.  
\label{pdffromcdf}
\eeq
The Mellin transform of $\p{m}{m}$ yields 
the moment 
$\aver{(\lamscrs{m}{(m)})^{q}}$.  \\

We note, for future reference, that the 
counterparts  
of Eq. (\ref{tracerhoq})  and 
Eqs. (\ref{psijl})--(\ref{cdfoflambdamax}) to 
general values of $m$ and $n \,(\geqslant m)$ 
are also available (Refs. \cite{bianchi} and 
\cite{vivo}, respectively). 
We proceed now to find the full set of 
PDFs $\{\p{k}{m}\}$ for different values  of $m$. 
Symbolic manipulation in 
{\it Mathematica 11} has been used to carry out all the calculations in what follows.  

\section{Two qubits: $m = n = 2$}
\label{sec:mn2}
A bipartite system of two qubits  
is a trivial case  as far as 
order statistics are concerned, since there 
are only two eigenvalues 
$\lamscrs{1}{(2)}$ and $\lamscrs{2}{(2)}
= 1- \lamscrs{1}{(2)}$.
Setting $m = 2$ in 
Eq. (\ref{pdflambdamin}), 
the  PDF of the smaller eigenvalue is  given, in this case, by 
\beq
\p{1}{2}= 6 (1- 2 x)^{2}\, \Theta(1 - 2 x).
\label{m2p1}
\eeq
Working out the steps outlined in
 Eqs. (\ref{psijl}) to (\ref{pdffromcdf}) with 
$m$ set equal to $2$, 
we obtain 
\beq
\p{2}{2}= 6 (1- 2 x)^{2} \left[ \Theta(1 - x) - \Theta(1 - 2 x) \right].  
\label{m2p2}
\eeq
Apart from the step functions, 
these are polynomials of order 
$m^{2}- 2 = 2$, with support in 
$[0, 1/2]$ and 
$[1/2, 1]$ respectively. 
The expression for $\p{2}{2}$ could 
have been written down from that 
for $\p{1}{2}$ in this case: 
Since 
$\Lambda_{1}^{(2)} + \Lambda_{2}^{(2)} 
= 1$,  it follows that   $\p{2}{2} = 
p_{1}^{(2)}(1-x)$.  \\ 

\begin{figure}[H]
\centering
\includegraphics[width=0.5\textwidth]{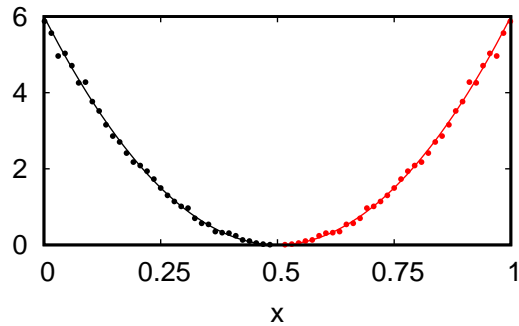}
\caption{PDFs  of the two eigenvalues 
in the case  $m=n=2$. In this figure and in all the figures that follow, the solid-line curves  correspond to the exact analytical expressions derived. 
The dots  represent numerical histograms obtained from an ensemble of over $10^{5}$ random pure states.}
\label{fig:mn2}
\end{figure}

In order to verify the expressions above  
and those  to be obtained 
for $\p{k}{m}$ 
in subsequent sections, we generate the histograms 
of $\lamscrs{k}{(m)}$ computed from an ensemble of 
random complex pure states of the composite system. 
We  consider a randomly chosen pure state of the full system $AB$ to be an $mn$-dimensional 
 column vector, with  the real and  imaginary parts of each element of the vector  drawn from a standard normal distribution.  
 The state is then normalized. 
The moments (and hence the cumulants) of the numerically generated histograms, obtained with  $1.001 \times 10^{5}$ random pure states, agree up  to the third decimal place  with those computed from the analytical expressions. We have also verified that this agreement improves on increasing the number of random pure states in the ensemble to $10^{6}$. These statements 
remain valid in all the cases to be considered in the sections that follow.   
Figure \ref{fig:mn2} shows that there is excellent agreement between the numerically generated histograms of $\lamscrs{1}{(2)}$ and $\lamscrs{2}{(2)}$, and  the analytical expressions of 
Eqs. (\ref{m2p1}) and 
(\ref{m2p2}) 
for the corresponding PDFs. We note that the histograms in this figure and in subsequent figures have been normalised in order to enable direct comparison with the calculated PDFs.

The average values of the two eigenvalues 
are $\aver{\lamscrs{1}{(2)}}= 1/8$ and 
$\aver{\lamscrs{2}{(2)}}= 7/8$, while the  variance is  $3/320$ in both cases. 
From the Mellin transforms of the PDFs 
in Eqs. (\ref{m2p1}) and 
(\ref{m2p2}) we get, for the $q^{\rm th}$ moments of the eigenvalues,  
\beq
\aver{(\lamscrs{1}{(2)})^{q}}= 
3! q! 2^{-q}\big/(q+3)!
\label{m2lambda1q}
\eeq
and
\beq
\aver{(\lamscrs{2}{(2)})^{q}}= 3! q! ( q^{2} + q + 2 - 2^{-q})\big/(q+3)!
\label{m2lambda2q}
\eeq
Their sum 
\beq
\aver{(\lamscrs{1}{(2)})^{q}}+ \aver{(\lamscrs{2}{(2)})^{q}} 
=3! q! \left( q^{2} + q + 2 \right)\big/(q+3)!
\label{m2tracerhoq}
\eeq
tallies with the corresponding expression 
for $\aver{\mathrm{Tr}\,(\rhoa^{(2)})^{q}}$ 
obtained from Eq.~(\ref{tracerhoq}).
 A similar 
  agreement 
  with the known result for 
 $\aver{\mathrm{Tr}\,(\rhoa^{(m)})^{q}}$
will serve as a 
further check on the correctness 
of all the PDFs to be derived 
in what follows.

\section{Two qutrits: $m = n =  3$}
\label{sec:mn3} 

A  bipartite system of two qutrits is the first non-trivial case owing to the existence of 
an intermediate eigenvalue $\lamscrs{2}{(3)}$ in between 
the smallest and largest eigenvalues $\lamscrs{1}{(3)}$ and 
$\lamscrs{3}{(3)}$. There is, however, a simple strategy to find the PDF of 
$\lamscrs{2}{(3)}$ in this instance. 
In brief, we first find 
 $\aver{(\lamscrs{1}{(3)})^{q}}$ and  
$\aver{(\lamscrs{3}{(3)})^{q}}$, 
and use these results along with that  
for $\aver{\mathrm{Tr}\,(\rhoa^{(3)})^{q}}$
to deduce  
$\aver{(\lamscrs{2}{(3)})^{q}}$. 
The inverse Mellin transform of the 
latter then yields the PDF 
$\p{2}{3}$. \\

When $m = n = 3$, 
Eq.~(\ref{pdflambdamin}) gives
for the  
PDF of the smallest eigenvalue 
the expression   
\beq
\p{1}{3}=24 (1- 3 x)^{7} \Theta (1- 3 x).
\label{m3p1}
\eeq
As before, working through the steps 
in Eqs.~(\ref{psijl}) to (\ref{pdffromcdf}) with 
$m = 3$, 
we obtain 
\beqa
\p{3}{3}&=& 24 (1-3 x)^7 \Theta (1 - 3 x)\
\nonumber \\
&-&48 (1-2 x)^3 (156 x^{4} - 165 x^{3} + 87 x^{2} - 15 x + 1) \Theta (1 - 2 x)
\nonumber \\
& +&24 (1-x)^3 (309 x^{4} - 354 x^{3} +132 x^{2} -18 x +1) \Theta (1-x).
\label{m3p3}
\eeqa
The Mellin transforms of the two PDFs 
in Eqs.~(\ref{m3p1}) and (\ref{m3p3}) 
yield, respectively,  
\beq
\aver{(\lamscrs{1}{(3)})^{q}} = 
\frac{8! q! 3^{-q}}{(q+8)!}
\label{m3lambda1q}
\eeq
and 
\beqa
\aver{(\lamscrs{3}{(3)})^{q} }&= &
\frac{8! q!}{2^{6}(q+8)!}
 \Big\{ 2^{4} 
(q^{4} + 2 q^{3} + 11 q^{2} + 10 q 
+ 12)\nonumber \\
& &- 2^{-q} (q^{4} + 14 q^{3} + 83 q^{2} + 70 q + 192) + 2^{6} 3^{-q} \Big\}.
\label{m3lambda3q}
\eeqa
On the other hand, setting $m = 3$ in 
Eq.~(\ref{tracerhoq}) gives 
\beq
\aver{\mathrm{Tr} (\rhoa^{(3)})^{q}}
= \frac{8! q!}{4 (q+8)!} 
( q^{4} + 2 q^{3} + 11 q^{2} + 10 q + 12).
\label{m3tracerhoq}
\eeq
An important point that we note here 
for  future reference is the following. 
After the ratio $q!/(q+8)!$ is simplified, 
the expression on the right-hand side of Eq. 
(\ref{m3tracerhoq}) is a rational function 
of $q$. There are no transcendental 
functions like $r^{-q}$ present in 
$\aver{\mathrm{Tr} (\rhoa^{(3)})^{q}}$. 
Hence its  inverse Mellin transform does 
not have any step functions of the form 
$\Theta(1-rx)$ where  $r > 1$. \\

\begin{figure}[H]
\centering
\includegraphics[width=0.8\textwidth]{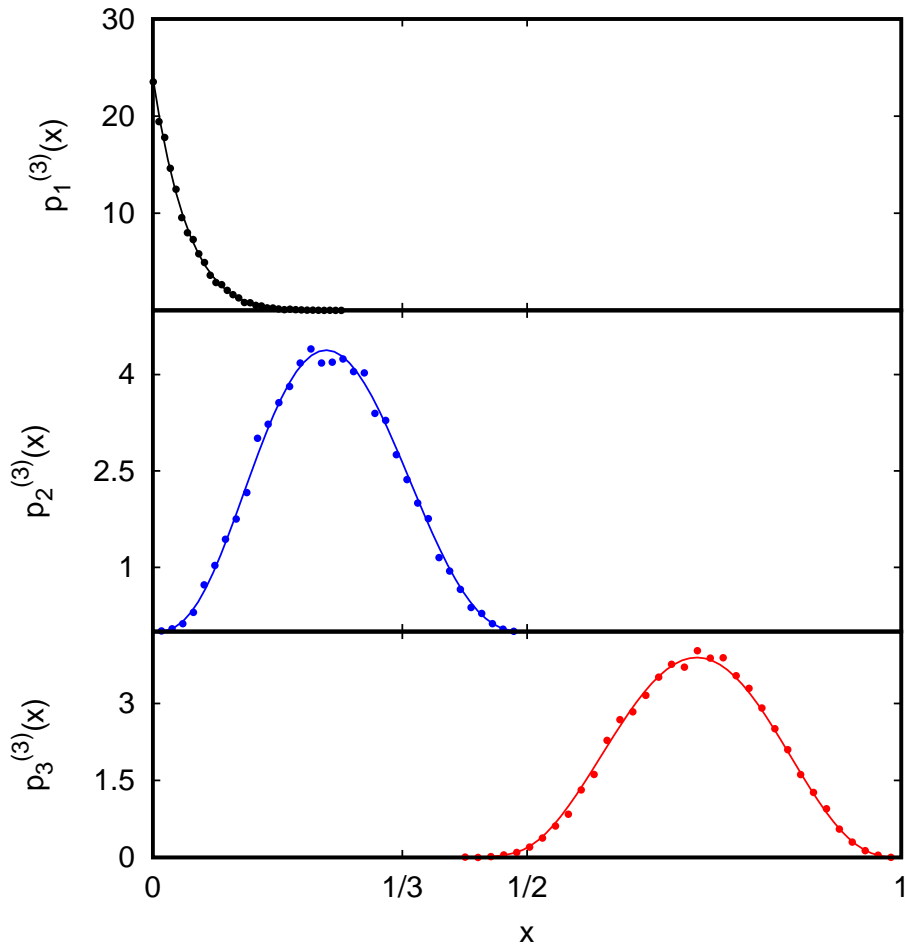}
\caption{PDFs of the ordered eigenvalues for $m=n=3$. Solid curves: analytical expressions; 
dots:  histograms from a 
Gaussian ensemble of 
random pure states.}
\label{fig:mn3}
\end{figure}

It follows from 
Eqs. (\ref{m3lambda1q})--(\ref{m3tracerhoq}) that 
 \beqa
 \aver{(\lamscrs{2}{(3)})^{q}} 
 & = & 
\aver{{\rm Tr} (\rhoa^{(3)})^{q}} - 
\aver{(\lamscrs{1}{(3)})^{q}} - \aver{(\lamscrs{3}{(3)})^{q}} \nonumber \\[6pt]
& = & 
\frac{8! q!}{2^{6} (q+8)!}
\Big\{2^{-q} (q^{4} + 14 q^{3} + 83 q^{2} + 70 q + 192) - 2^{7} 3^{-q}\Big\}. 
\label{m3lambda2q}
\eeqa
Inverting the Mellin transform, we
obtain  for the 
PDF of the middle eigenvalue the explicit expression  
\beqa
\p{2}{3}&=& 48 (1- 2 x)^{3} (156 x^{4} -165 x^{3} + 87 x^{2} -15 x+1) 
\Theta(1-2 x)\nonumber \\[4pt]
&&-48 (1-3 x)^{7} \Theta(1 - 3 x).
\label{m3p2}
\eeqa
Figure \ref{fig:mn3} again shows that the three PDFs 
$\p{k}{3}, k = 1,2,3$ are in excellent agreement with the numerically generated histograms. 
As the PDFs are essentially polynomials 
with compact support in ranges whose end-points are rational numbers, all the moments 
of these PDFs (and hence their cumulants) are rational numbers. (This feature remains valid for 
all values of $m$ and $n$.) 
Table \ref{tab:1} lists the values of the basic 
descriptors of the distributions concerned 
in terms of the corresponding 
cumulants $\kappa_{r}$: the mean 
$\kappa_{1}$, the variance $\kappa_{2}$, 
the skewness 
$\kappa_{3}^{2}/\kappa_{2}^{3}$, and the excess of kurtosis 
$\kappa_{4}/\kappa_{2}^{2}$. 
\begin{table}[H]
\begin{center}
\caption{\label{tab:1} Values of the 
descriptors  corresponding to the PDFs
$\p{k}{3}$.}
\begin{tabular}{lccc}
\hline
\hline
\phantom{xxxx}  &
$\p{1}{3}$ & $\p{2}{3}$& $\p{3}{3}$  \\
\hline
\hline
&&&\\
$\kappa_{1}$  & 
$\frac{1}{27}$& 
$\frac{103}{432}$ &  
$\frac{313}{432}$  \\
&&&\\
$\kappa_{2}$ & 
$\frac{4}{3645}$ &
$\frac{6499}{933120}$ & 
$\frac{8179}{933120}$ \\
&&&\\
$\kappa_{3}^{2}/\kappa_{2}^{3}$ & 
$\frac{245}{121}$ &
$\frac{241916407220}{33214290609379}$ & 
$\frac{80059327220}{66204269040019}$\\ 
&&&\\
$\kappa_{4}/\kappa_{2}^{2}$  & 
$\frac{201}{88}$ & 
$-\frac{248949138}{464607011}$ & 
$-\frac{387186258}{735856451}$ \\
&&&\\
\hline
\hline
\end{tabular}
\end{center}
\end{table}

\section{The case  $m = n = 4$}
\label{sec:mn4} 
It is clear that the simple argument used in the case 
$ m = 3$ is no longer applicable when there is more than 
a single intermediate eigenvalue, i.e., for any $m \geqslant 4$.  
There is, however, a way to deduce 
the PDF $\p{k}{m}$ 
for every one of these eigenvalues. The case 
$m = 4$ serves as the simplest illustration of this 
method. As before, 
we  start with  the 
PDF of the smallest eigenvalue, obtained by setting $m = n = 4$ in Eq. (\ref{pdflambdamin}). 
We have  
\beq
\p{1}{4} =60 (1-4 x)^{14} \Theta(1-4 x).
\label{m4p1} 
\eeq
Next, we find the explicit expression 
for the PDF  $\p{4}{4}$ of the largest 
eigenvalue 
from Eqs.~(\ref{psijl}) -- (\ref{pdffromcdf})
for  $m = 4$.  
It is convenient to introduce the notation 
$A_{j}^{(4,4)}(x)$ ($j=1,2,3,4$) for 
the polynomial 
that is the coefficient
of $\Theta(1- j x)$     
in this expression.  
(The superscripts indicate the values of $m$ and $n$.) 
We then find that  
\beqa
\p{4}{4}&= &-A_{4}^{(4,4)}(x) \Theta(1-4 x) + A_{3}^{(4,4)}(x) 
\Theta(1-3 x)\nonumber \\[4pt]
&& - A_{2}^{(4,4)}(x) \Theta(1-2 x)+ A_{1}^{(4,4)}(x) \Theta(1-x),
\label{m4p4}
\eeqa
where 
\beqa
A_{4}^{(4,4)}(x)&=&60 (1-4 x)^{14},
\nonumber\\[4pt]
A_{3}^{(4,4)}(x)&=&60 (1-3 x)^{8} \big(3 - 96 x + 1308 x^{2} - 6128 x^{3} + 
29818 x^{4}\nonumber \\
&& \hspace*{2 em}  - 70160 x^{5} 
+ 67812 x^{6}\big),\nonumber \\[4pt]
A_{2}^{(4,4)}(x)&=&30 (1-2 x)^{6} \big(6 - 264 x + 5208 x^{2} - 45920 x^{3} + 229936 x^{4}
\nonumber \\
&& \hspace*{2 em}  - 859040 x^{5} + 2706592 x^{6} - 5570528 x^7+5517256 x^8\big),
\nonumber \\[4pt]
A_{1}^{(4,4)}(x)&=&60(1-x)^{8}
\big(1-48 x+ 1044 x^2 -9904 x^3
+44934 x^4 \nonumber \\
&& \hspace*{2 em}   
- 94128 x^5
+73116 x^6\big).
\label{m4Ai}
\eeqa
We observe  that 
\beq
-A_{4}^{(4,4)}(x) + A_{3}^{(4,4)}(x) 
-A_{2}^{(4,4)}(x) + A_{1}^{(4,4)}(x) 
=  0,
\label{m4Aisum}
\eeq
 ensuring that $\p{4}{4}$ vanishes identically for $x < 1/4$, (As we know, its support 
 is $[1/4, 1]$). Next, setting $m = 4$ in 
 Eq.~(\ref{tracerhoq}), we get 
\beq
\aver{\mathrm{Tr} (\rhoa^{(4)})^{q}}
=\frac{15! q!}{36 (q+15)!} 
\big(144 + 
156 q + 184 q^2 + 57 q^3 + 31 q^4 + 
3 q^5 + q^6\big). 
\label{m4tracerhoq}
\eeq
Once again, we  note  that 
the expression on the right-hand side of 
Eq. (\ref{m4tracerhoq}) is a rational function 
of $q$ (after the ratio $q!/(q+15)!$  
is simplified). Hence, by the property 
(a) of Mellin transforms noted in 
Section \ref{sec:preliminaries},  the step functions $\Theta(1-4x),  
 \Theta(1-3x)$    and $\Theta(1-2x)$ cannot appear 
 in   its inverse Mellin transform 
 $\sum_{k=1}^{4}\p{k}{4}$. 
 The coefficients of these step functions must therefore vanish 
 identically when the individual PDFs are added up.
 Note also that 
 $\p{1}{4}  = A_{4}^{(4,4)}(x)\Theta(1-4x)$. 
These facts lead us naturally to the ansatz that $\p{2}{4}$ and $\p{3}{4}$  must have the  forms 
\beq
\p{2}{4}=-c_{1} A_{4}^{(4,4)}(x) \Theta(1-4 x) + c_{2} A_{3}^{(4,4)}(x) \Theta(1-3 x),
\label{m4ansatz}
\eeq
\beqa
\p{3}{4}= &c_{1} A_{4}^{(4,4)}(x) \Theta(1-4 x) 
- (c_{2}+1) A_{3}^{(4,4)}(x) \Theta(1-3 x) 
\nonumber \\[3pt]
& + A_{2}^{(4,4)}(x) \Theta(1-2 x),
\label{}
\eeqa
where $c_{1}$ and $c_{2}$ are constants.  
 They are determined from the normalization 
 (to unity)  of 
$\p{2}{4}$ and $\p{3}{4}$ in the ranges 
$[0, 1/3]$ and $[0, 1/2]$, respectively.  
Using the fact that 
$\int_{0}^{1/4} A_{4}^{(4,4)}(x) \rmd x = 1,\, 
 \int_{0}^{1/3} A_{3}^{(4,4)}(x) \rmd x 
= 4, \,
\int_{0}^{1/2} A_{2}^{(4,4)}(x) \rmd x = 6$ and 
$\int_{0}^{1} A_{1}^{(4,4)}(x) \rmd x = 4$, 
 we get $c_{1} = 3,\,c_{2}= 1$. Hence
\beqa
\p{2}{4}&=& -3 A_{4}^{(4,4)}(x) \Theta(1-4 x) + A_{3}^{(4,4)}(x) \Theta(1-3 x),\\[4pt]
\p{3}{4}&= & 3 A_{4}^{(4,4)}(x) \Theta(1-4 x) - 2 A_{3}^{(4,4)}(x) \Theta(1-3 x)\nonumber
\\[2pt]
&& + A_{2}^{(4,4)}(x) \Theta(1-2 x).
\label{m4p2p3}
\eeqa
We observe from the foregoing (and from all the cases to be considered in the sequel)  
that the constants multiplying the coefficients 
$A_{j}^{(m,n)}(x) $ for a given $j$ in the different PDFs $\p{k}{m}$ 
($m-j+1 \leqslant k 
\leqslant m$) are the binomial 
coefficients ${j-1}\choose{m-k}$ with alternating signs. This fact 
also guarantees that the step functions 
(other than $\Theta (1-x)$) do not appear 
in $\sum_{k=1}^{4}\p{k}{4}$, and more generally 
in  $\sum_{k=1}^{m}\p{k}{m}$.  \\

\begin{figure}[H]
\centering
\includegraphics[width=0.8\textwidth]{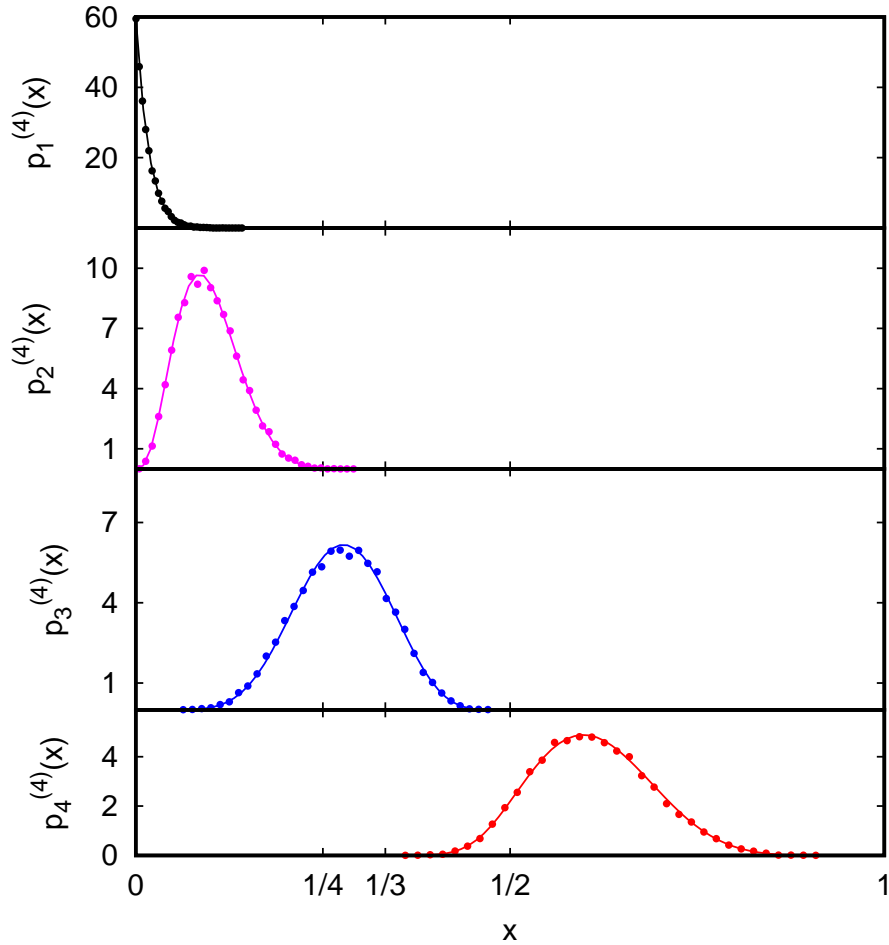}
\caption{PDFs  of the ordered eigenvalues for $m=n=4$. 
Solid curves: analytical expressions; dots: histograms from a Gaussian  ensemble of random pure states.}
\label{fig:mn4}
\end{figure}

The four PDFs $\p{k}{4},  \,1 
\leqslant k  \leqslant 4$, are plotted  
 in Figure \ref{fig:mn4}. Once again, there 
 is excellent agreement with the 
  numerically generated histograms of the ordered eigenvalues.  The mean values of the 
 four  eigenvalues are found to be
  \beq
  \aver{\lamscrs{1}{(4)}} = \textstyle{\frac{1}{64}}, \;
  \aver{\lamscrs{2}{(4)}} = \textstyle{\frac{13727} {139968 }}, \;
  \aver{\lamscrs{3}{(4)}} = \textstyle{\frac{617057}{2239488}},\;
  \aver{\lamscrs{4}{(4)}} = \textstyle{\frac{1367807}{2239488}}.  
  \label{m4meanvalues}
  \eeq
  The higher cumulants can also be calculated, and they are all rational numbers.
  We have also verified that the sum of the 
$q^{\rm th}$ moments of the eigenvalues tallies with the known expression 
for $\aver{\mathrm{Tr} (\rhoa^{(4)})^{q}}$.

\section{Other cases}
\label{sec:others}
\begin{figure}[H]
\centering
\includegraphics[width=0.8\textwidth]{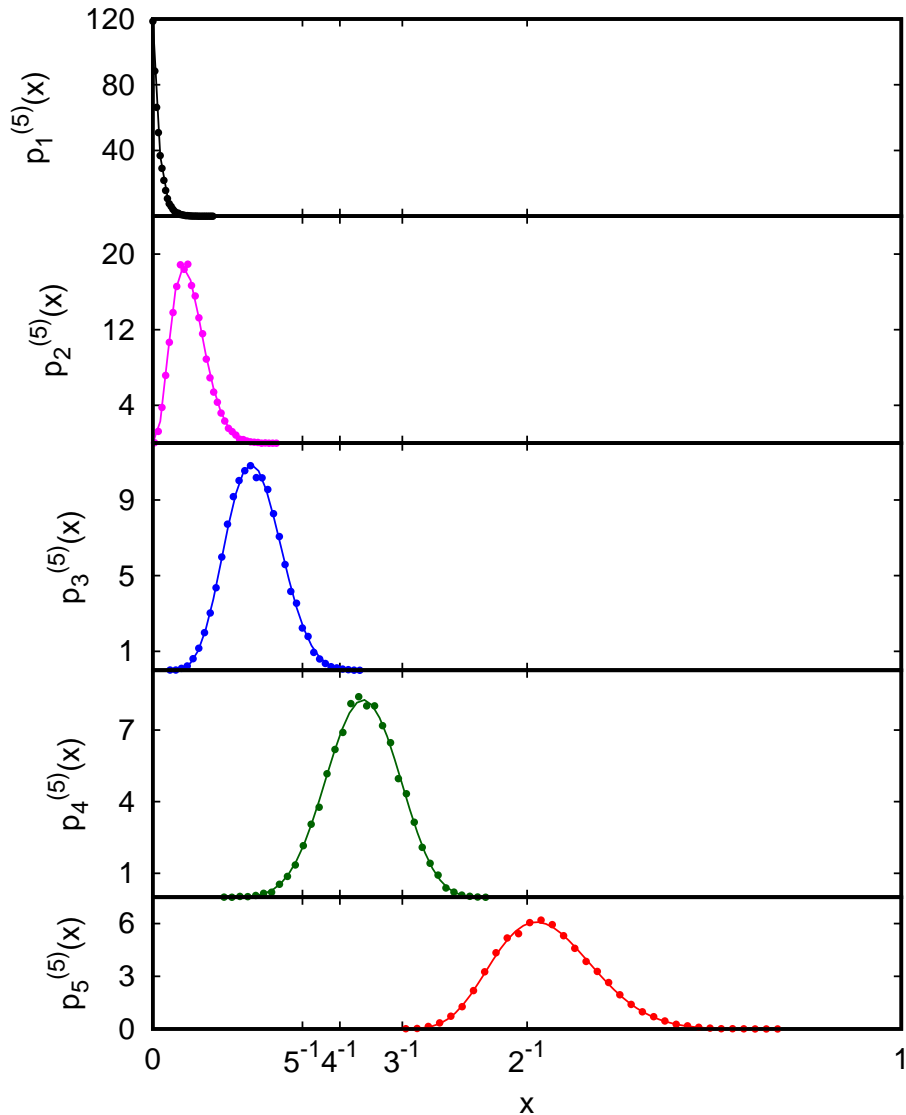}
\caption{PDfs  of  the ordered eigenvalues for $m=n=5$.}
\label{fig:mn5}
\end{figure}

As  further checks  of the method used, 
we have carried out  similar 
calculations to determine 
the PDFs of the ordered eigenvalues in the cases 
$m = n = 5, 6$ and $7$, respectively.  The algebraic 
expressions become considerably more 
lengthy as $m$ increases. 
The expressions for the PDFs when  
$m=n = 5$ are given in  Appendix A, and these
expressions agree very well with the numerically generated histograms, as shown  in Figure \ref{fig:mn5}. As already pointed out, we find that the constants multiplying the coefficient functions 
$A_{j}^{(5,5)}(x)$ 
are appropriate binomial coefficients with alternating signs.
\noindent
The expressions obtained for the PDFs in the case $m = n = 6$ are also recorded  in Appendix A. Once again, we have also verified that there is very good  agreement between the analytical expressions for the PDFs and the numerically generated histograms.  Similarly, the expressions 
for $m = n = 7$ are also 
precisely along expected lines, and will not be given here. \\ 

\begin{figure}[H]
\centering
\includegraphics[width=0.8\textwidth]{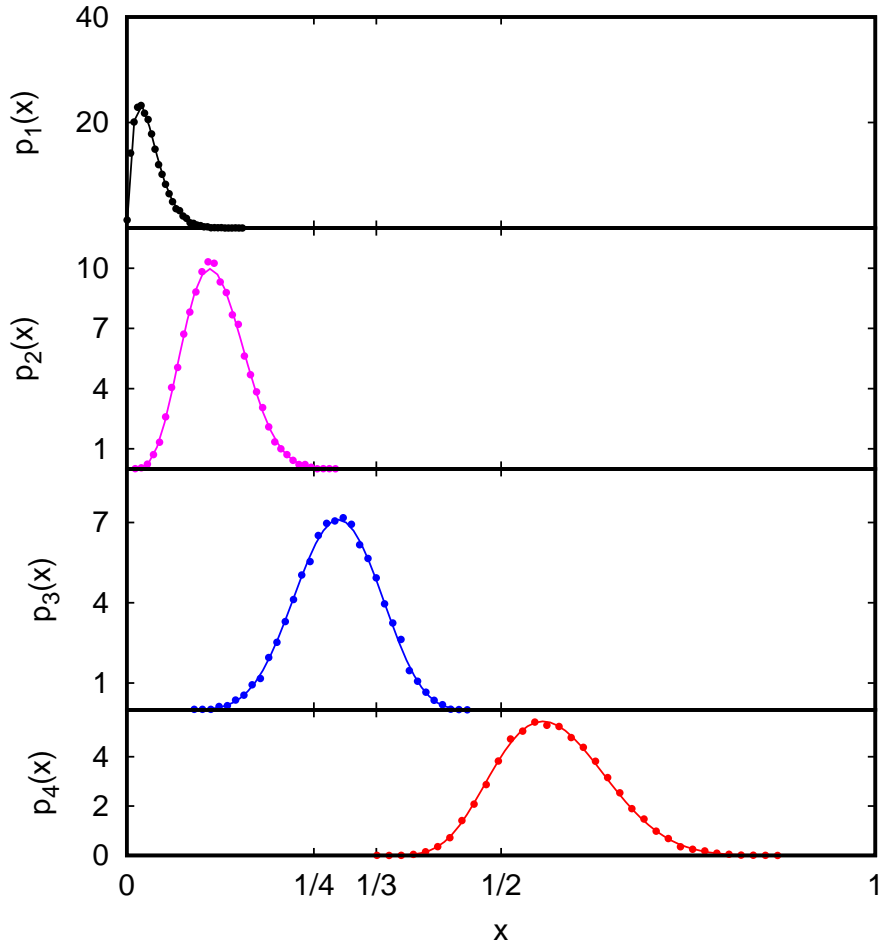}
\caption{PDFs of the ordered eigenvalues 
when  $m=4$, $n=5$.}
\label{fig:m4n5}
\end{figure}

Finally, in order to show 
 that our method  works even 
when  $m \neq n$, we have found  the analytical expressions for  the PDFs when  $m=4$ 
and  $n=5$. We must now take into account the fact that  the index  $\alpha = 1$ in this case, and use the corresponding generalizations of Eqs. (\ref{tracerhoq})-- 
(\ref{pdffromcdf}). The details are given in Appendix B.
Once again, the plots of  the calculated PDFs 
are in complete  agreement with the numerical histograms, as shown in Figure \ref{fig:m4n5}.
Table \ref{tab:2} lists  
the averages $\aver{\Lambda_{k}^{(m)}}$ 
for the three cases considered 
in this section. 
\begin{table}
\begin{center}
\caption{\label{tab:2} Mean values 
of the ordered eigenvalues 
for $m = 4,  n = 5$;  
 $m=n=5$;  and $m=n=6$.} 
\begin{tabular}{cccc}
\hline
\hline
Mean & $m = 4$, $n = 5$ & 
$m=n=5$ & $m=n=6$ \\
\hline
\hline
&&&\\
$\aver{\Lambda_{1}^{(m)}}$ & 
$\frac{125}{4096}$ &
$\frac{1}{125}$ & $\frac{1}{216}$ \\ 
&&&\\
$\aver{\Lambda_{2}^{(m)}}$ & 
 $\frac{3188009}{26873856}$ &
$\frac{813587}{16384000}$ & $\frac{301301927}{10546875000}$\\
&&&\\
$\aver{\Lambda_{3}^{(m)}}$ & 
$\frac{7552985}{26873856}$ &
$\frac{1182578796887}{8707129344000}$ & $\frac{873543307049548733}{11324620800000000000}$ \\
&&&\\
$\aver{\Lambda_{4}^{(m)}}$ 
& $\frac{15312737}{26873856}$&
 $\frac{2440637328617}{8707129344000}$ & $\frac{75602489231060183229976073}{487487792008396800000000000}$\\
&&&\\
$\aver{\Lambda_{5}^{(m)}}$ & &$\frac{4581882694877}{8707129344000}$ & $\frac{132969850997476498208010743}{487487792008396800000000000}$  \\
&&&\\
$\aver{\Lambda_{6}^{(m)}}$ & & &$\frac{225128892964655720357665283}{487487792008396800000000000}$  \\
&&&\\
\hline
\hline
\end{tabular}
\end{center}
\end{table}

\section{Solution for general $m$ and $n$}
\label{sec:generalmn}
We now proceed to  the exact formal expression 
for the PDF $p_{k}^{(m,n)}(x)$ of the  
$k^{\rm th}$ eigenvalue order 
statistic $\lamscrs{k}{(m)}, \,1 \leqslant k \leqslant m$, 
for general values of the subsystem dimensions $m$ and 
$n \geqslant m$. The procedure followed is the same 
as that for the case $m = n$. As already mentioned, the 
counterparts of Eqs. (\ref{tracerhoq}) \cite{bianchi}
and 
Eqs.~(\ref{psijl})--(\ref{cdfoflambdamax}) \cite{vivo} 
for the case $n \geqslant m$ are now required. 
The pattern in the structure of the PDFs found in the 
foregoing sections aids us considerably  in deducing  the 
structure for general $m$ and $n$.  
We obtain, finally, 
\beq
\p{k}{m,n} 
= \frac{1}{\mathcal{N}} \sum\limits_{j=m-k+1}^{m} (-1)^{m-k+j+1} 
\textstyle{{{j-1}\choose{m-k}}} \, A_{j}^{(m,n)}(x) \, 
\Theta(1-j x), 
\label{pdfgeneralmn}
\eeq
where   $A_{j}^{(m,n)}(x)$ is a polynomial in $x$ of order $mn-1$, to be specified in Eqs. (\ref{Amngeneral})--(\ref{etadefn}) below. 
The  constant $\mathcal{N}$ is determined by normalizing 
$\p{k}{m,n}$ to unity in the 
sub-interval $I_{k}^{(m)}$ of the unit 
interval in which it has 
a support. \\

Let  $\boldsymbol{\sigma} =
(\sigma_{0}, \sigma_{1}, \ldots, \sigma_{m-1})$ be a permutation of the sequence $\{ 0,1,\ldots,m-1)$,  with 
$\mathrm{sgn} \,\boldsymbol{\sigma}=\pm 1$ 
depending on whether  $\boldsymbol{\sigma}$ is an 
even or odd permutation of the natural 
order,  and let  $S$ denote the set of all permutations 
$\boldsymbol{\sigma}$.  Setting 
$a=m-j$ where  $ 1 \leqslant j \leqslant m$, 
we have 
\beqa
A_{j}^{(m,n)}(x)&=& - (m n-1)! \prod_{j=0}^{m-1}
\frac{(j+1)! j! (n-m+j)!}
{(n-1-j)! (m-j)! (n+j)!} \times \nonumber \\
&& (d/d x)\Big[ x^{m n-1} \sum \limits_{\boldsymbol{\sigma} \in S} (\mathrm{sgn}\,\boldsymbol{\sigma}) \Big\{ \sum \limits_{k_{1}=0}^{m-a} \,\sum\limits_{k_{2}=k_{1}+1}^{m-a+1} \cdots \sum\limits_{k_{a}=k_{a-1}+1}^{m-1} \times
\nonumber \\
&& \prod_{i=0}^{m-1} \,\sum\limits_{\ell_{i}=0}^{n-m+i+\sigma_{i}} \xi \left(x^{-1} - j \right)^{\eta}/\eta ! \Big\} \Big], 
\label{Amngeneral}
\eeqa
where 
\beq
\xi =  {\textstyle{{n-m+i+\sigma_{i}}\choose{\ell_{i}}}} \,\ell_{i}! \, \bigg(1-\sum\limits_{b=1}^{a} \delta_{i,k_{b}}\bigg) + \sum\limits_{c=1}^{a} \delta_{i,k_{c}} \,\delta_{\ell_{i},0} \,(n-m+i+\sigma_{i})! 
\label{xidefn}
\eeq
and
\beq
\eta= m-1+\sum\limits_{i=0}^{m-1} 
\Big\{ \bigg(1-\sum\limits_{b=1}^{a} \delta_{i,k_{b}}\bigg) \ell_{i} + \sum\limits_{c=1}^{a} \delta_{i,k_{c}}\, (n-m+i+\sigma_{i})
\Big\}.
\label{etadefn}
\eeq
It is evident that the general solution for the 
PDF $\p{k}{m,n}$, while exact and explicit, is
algebraically  quite involved. This fact 
further corroborates the usefulness of 
displaying in detail the results for several 
small values of $m$, as has been done in the foregoing. 
 
 To summarize: 
 We have  obtained  the 
probability density functions of the 
eigenvalue order statistics 
$\lamscrs{k}{(m)}\,(1 \leqslant k \leqslant m)$ 
corresponding to the reduced density matrices for  
a Gaussian  ensemble of random 
complex pure states of a 
bipartite system, where $m$ is the smaller 
subsystem dimensionality. 
The  PDF $\p{k}{m,n}$ of the ordered 
 eigenvalue $\lamscrs{k}{(m)}$
is a linear combination of unit step functions 
$\Theta(1-mx), \ldots, 
\Theta\big(1-(m+1-k)x\big)$, each  multiplied by a polynomial
of order $m^{2}-2$ when $n = m$, and of order 
$mn - 2$ when $n > m$.  
The support of $\p{k}{m,n}$ is $[0, 1/(m-k+1)]$ 
for $1 \leqslant k \leqslant m-1$, and 
$[1/m, 1]$ for $k = m$.  
In all the cases considered, the 
 analytic expressions obtained for the  PDFs
 are in excellent  agreement  with the numerically generated histograms of the eigenvalues 
 concerned. As further corroboration, 
 we also find that, in every  case, the Mellin transform  of the 
 sum of the $q^{\rm th}$ moments of these PDFs matches 
 the known expression\cite{bianchi}  for    
 $\aver{\mathrm{Tr}\big(\rhoa^{(m)}\big)^{q}}$.  \\

Based on the explicit analytic solutions 
in the cases $m = 3,4,5, 6$ and $7$, we 
deduce the following general properties.
When  $m=n$,  the PDF $\p{1}{m}$ of the smallest eigenvalue decreases monotonically from 
the value $p_{1}^{(m)}(0)
= m(m^{2}-1)$ to 
the value $p_{1}^{(m)}(1/m)= 0$ as 
$x$ increases from $0$ to $1/m$.
When $m < n$, however, 
$p_{1}^{(m)}(0)= 0$. 
 Reverting to $m=n$, every $\p{k}{m}$ (where  $2\leqslant k \leqslant m-1$) vanishes 
like $x^{k^{2}-1}$ as $x \rightarrow 0$. In the limit $x \rightarrow
1/(m-1)$, $\p{2}{m}$ vanishes like 
$\big(1-(m-1)x\big)^{m^{2}-2 m}$.
The PDF $\p{k}{m}$ for both $k=2+j$ and $k = m-j$ ($j=1,2,\dots, \lfloor \frac{m}{2} \rfloor - 1 $) vanishes like 
$\big(1- (m+1-k)x\big)^{r}$ 
as $x \rightarrow 1/(m+1-k)$, 
where 
$r = m^{2}-2 m - 2 \sum_{i=1}^{j} (m- 2 i -1)$.  
The PDF  $\p{m}{m}$ of the largest eigenvalue 
$\lamscrs{m}{(m)}$     vanishes like $(1-mx)^{m^{2}-2}$ as 
$x \rightarrow 1/m$ from above, and like 
$(1-x)^{m^{2}-2 m}$ as $x \rightarrow 1$ from below.

\ack
We would like to thank Arul Lakshminarayan for drawing  our attention to Ref. \cite{forrester}. This work was supported in part by a seed grant from IIT Madras to the Centre
for Quantum Information Theory of Matter and Spacetime, under the IoE-CoE scheme.

\providecommand{\newblock}{}

\appendix
\section*{Appendix A}
\setcounter{section}{1}
We first present the analytic expressions of the PDFs of the ordered eigenvalues $\lbrace \lamscrs{k}{(5)} \rbrace$ for $m=n=5$. The PDFs $\p{k}{5}$ ($k=1,2,\dots,5$) are written in terms of the functions $A_{j}^{(5,5)}(x)$ ($j=1,2,\dots,5$) that  are the coefficients of the respective step functions $\Theta(1-j x)$. These coefficient functions are as listed below.
\beqa
A_{5}^{(5,5)}(x)&=&120 \,(1-5 x)^{23},
\nonumber \\
A_{4}^{(5,5)}(x)&=&240 \,(1-4 x)^{15} \left(2-110 x+2690 x^2-20600 x^3+304595 x^4 \right.\nonumber\\
&& \left. -1558835 x^5 +4852905 x^6-10365975 x^7+11082660 x^8\right),\nonumber\\
A_{3}^{(5,5)}(x)&=&720 \,(1 - 3 x)^{11} \left(1-82 x+3124 x^2-55528 x^3+656656 x^4 
\right.\nonumber\\
&& -6833200 x^5 +60965520 x^6-390601200 x^7+1733312295 x^8\nonumber\\
&& -5065359970 x^9 +10140970180 x^{10}-13794793180 x^{11}\nonumber\\
&& \left. +11635970460 x^{12} \right),\nonumber\\
A_{2}^{(5,5)}(x)&= & 240 \,(1-2 x)^{11} \left(2-186 x+8118 x^2-167464 x^3+2021877 x^4
\right.\nonumber\\
&& -18428355 x^5 +161532525 x^6-1281331755 x^7+7805513430 x^8
\nonumber \\
&& -33503168380 x^9 + 94797708060 x^{10}-158275026540 x^{11}\nonumber\\
&& \left. +119326518320 x^{12} \right),
\nonumber\\
A_{1}^{(5,5)}(x)&=&120 \,(1-x)^{15} \left(1-100 x+4720 x^2-104200 x^3+1215160 x^4
\right.\nonumber\\
&& \left. -7812880 x^5+27619440 x^6-49896300 x^7+35838555 x^8 \right).
\eeqa
In  terms of these polynomials, the 
PDFs $\{\p{k}{5}\}$ are  found to be  
\beqa
\p{1}{5}&= &A_{5}^{(5,5)}(x) \Theta(1-5 x),\\[6pt]
\p{2}{5}&= &- 4 A_{5}^{(5,5)}(x) \Theta(1-5 x) + A_{4}^{(5,5)}(x) \Theta(1-4 x),\\[6pt]
\nonumber \p{3}{5}&=& 6 A_{5}^{(5,5)}(x) \Theta(1-5 x) - 3 A_{4}^{(5,5)}(x) 
\Theta(1-4 x)\\[4pt]
&& + A_{3}^{(5,5)}(x) \Theta(1-3 x),\\[6pt]
\nonumber \p{4}{5}&=& - 4 A_{5}^{(5,5)}(x) \Theta(1-5 x) + 3 A_{4}^{(5,5)}(x) 
\Theta(1-4 x)\\[4pt]
&& - 2 A_{3}^{(5,5)}(x) \Theta(1-3 x) + A_{2}^{(5,5)}(x)  \Theta(1-2 x),\\[6pt]
\nonumber \p{5}{5}&=& A_{5}^{(5,5)}(x) \Theta(1-5 x) - A_{4}^{(5,5)}(x) 
\Theta(1-4 x)\\[4pt]
\nonumber &&+ A_{3}^{(5,5)}(x) \Theta(1-3 x) - A_{2}^{(5,5)}(x)  \Theta(1-2 x)\\[4pt]
&& + A_{1}^{(5,5)}(x) \Theta(1-x).
\label{eqn:probsM5}
\eeqa
As  before, the constants multiplying 
$A_{j}^{(m,n)}(x) $ for a given $j$ in different 
PDFs $\p{k}{m}$  ($1 \leqslant k 
\leqslant m$) are the binomial coefficients 
${j-1}\choose{m-k}$ with alternating signs.  \\

We also report the analytic expressions of the PDFs of the ordered eigenvalues $\lbrace \lamscrs{k}{(6)} \rbrace$ for $m=n=6$. As before, the PDFs $\p{k}{6}$ ($k=1,2,\dots,6$) are written in terms of the functions $A_{j}^{(6,6)}(x)$ ($j=1,2,\dots,6$), which are listed below.
\beqa
A_{6}^{(6,6)}(x)&=&210 \,(1-6 x)^{34},
\nonumber \\
A_{5}^{(6,6)}(x)&=&210 \,(1-5 x)^{24} \left(5 - 420 x + 16080 x^2 - 160680 x^3 + 6469230 x^4 \right.\nonumber\\
&& \left. - 40658112 x^5 + 261366628 x^6 - 1595391672 x^7 + 5683720173 x^8\right.\nonumber\\
&& \left. - 11348219292 x^9 + 11273058660 x^{10}\right), \nonumber\\
A_{4}^{(6,6)}(x)&=&420 \,(1-4 x)^{18} \left(5 - 660 x + 41220 x^2 - 1199200 x^3 \right.\nonumber\\
&& \left. + 26188080 x^4 - 541359744 x^5 + 9132924768 x^6 - 109228380096 x^7 \right.\nonumber\\
&& \left. + 969229595664 x^8 - 6384003186176 x^9 + 35245566675264 x^{10} \right.\nonumber\\
&& \left. - 168039178157376 x^{11} + 674535601042864 x^{12} - 2058660341189376 x^{13} \right.\nonumber\\
&& \left. + 4315240551175584 x^{14} - 5476040960131392 x^{15} + 3527358922055856 x^{16} \right), \nonumber\\
A_{3}^{(6,6)}(x)&=&420 \,(1 - 3 x)^{16} \left(5 - 780 x + 58140 x^2 - 2125680 x^3 \right.\nonumber\\
&& \left. + 50119740 x^4 - 1004003136 x^5 + 19201278456 x^6 - 311887564848 x^7 \right.\nonumber\\
&& \left. + 3949780543830 x^8 - 38228455420056 x^9 + 283595869865088 x^{10} \right.\nonumber\\
&& \left. - 1648254166845840 x^{11} + 7876735652844396 x^{12} - 32847798731822496 x^{13} \right.\nonumber\\
&& \left. + 122296710227124168 x^{14} - 385032778740807120 x^{15} + 925473909342876741 x^{16} \right.\nonumber\\
&& \left. - 1465716247992173916 x^{17} + 1154580059692232388 x^{18} \right), \nonumber\\
A_{2}^{(6,6)}(x)&= & 210 \,(1-2 x)^{18} \left(5 - 840 x + 67680 x^2 - 2641760 x^3 + 60875040 x^4 \right.\nonumber\\
&& \left. - 1041040128 x^5 + 17346863424 x^6 - 289429058688 x^7 + 4135247214912 x^8 \right.\nonumber\\
&& \left. - 46316923954048 x^9 + 395768314525056 x^{10} - 2538512485868160 x^{11} \right.\nonumber\\
&& \left. + 11970268586045536 x^{12} - 40164721924654464 x^{13} + 90652008902870976 x^{14} \right.\nonumber\\
&& \left. - 123384033397219200 x^{15} + 76712186285087664 x^{16} \right), \nonumber\\
A_{1}^{(6,6)}(x)&=&210 \,(1-x)^{24} \left(1 - 180 x + 15600 x^2 - 657000 x^3 + 15307350 x^4 \right.\nonumber\\
&& \left. - 210235104 x^5 + 1758025460 x^6 - 8979492600 x^7 + 27172972425 x^8 \right.\nonumber\\
&& \left. - 44490525420 x^9 + 30241971348 x^{10}\right). \nonumber\\
\eeqa
In  terms of these polynomials, the 
PDFs $\{\p{k}{6}\}$ are found to be  
\beqa
\p{1}{6}&= &A_{6}^{(6,6)}(x) \Theta(1-6 x),\\[6pt]
\p{2}{6}&= &- 5 A_{6}^{(6,6)}(x) \Theta(1-6 x) + A_{5}^{(6,6)}(x) \Theta(1-5 x),\\[6pt]
\nonumber \p{3}{6}&=& 10 A_{6}^{(6,6)}(x) \Theta(1-6 x) - 4 A_{5}^{(6,6)}(x) 
\Theta(1-5 x)\\[4pt]
&& + A_{4}^{(6,6)}(x) \Theta(1-4 x),\\[6pt]
\nonumber \p{4}{6}&=& - 10 A_{6}^{(6,6)}(x) \Theta(1-6 x) + 6 A_{5}^{(6,6)}(x) 
\Theta(1-5 x)\\[4pt]
&& - 3 A_{4}^{(6,6)}(x) \Theta(1-4 x) + A_{3}^{(6,6)}(x)  \Theta(1-3 x),\\[6pt]
\nonumber \p{5}{6}&=& 5 A_{6}^{(6,6)}(x) \Theta(1-6 x) - 4 A_{5}^{(6,6)}(x) 
\Theta(1-5 x)\\[4pt]
\nonumber &&+ 3 A_{4}^{(6,6)}(x) \Theta(1-4 x) - 2 A_{3}^{(6,6)}(x)  \Theta(1-3 x)\\[4pt]
&& + A_{2}^{(6,6)}(x) \Theta(1- 2 x). \\[4pt]
\nonumber \p{6}{6}&=& - A_{6}^{(6,6)}(x) \Theta(1-6 x) + A_{5}^{(6,6)}(x) \Theta(1-5 x) - A_{4}^{(6,6)}(x) 
\Theta(1-4 x)\\[4pt]
\nonumber &&+ A_{3}^{(6,6)}(x) \Theta(1-3 x) - A_{2}^{(6,6)}(x)  \Theta(1-2 x)\\[4pt]
&& + A_{1}^{(6,6)}(x) \Theta(1-x).
\label{eqn:probsM6}
\eeqa
As already pointed out, we find that the constants multiplying the coefficient functions 
$A_{j}^{(6,6)}(x)$ 
are appropriate binomial coefficients with alternating signs.

\section*{Appendix B}
\setcounter{section}{2}
We consider the case  $m=4$ 
and  $n=5$, in order to show that   
our method  works even 
when  $m \neq n$.
We must  take into account the fact that  the index  $\alpha$, 
defined in Eq. (\ref{alphadefn}), is now equal to 
$1$. Using the corresponding generalizations of Eqs. (\ref{tracerhoq})-- 
(\ref{pdffromcdf}), we find the explicit expression 
for the PDF  $\p{4}{4,5}$ of the largest 
eigenvalue~\cite{vivo}. (In the general case 
$m < n$, this PDF is found to be a linear 
combination of 
polynomials of order $mn - 2$ 
multiplied by appropriate step functions.)
We use the notation 
$A_{j}^{(4,5)}(x)$ ($j=1,2,3,4$) for 
the polynomial 
that is the coefficient
of $\Theta(1- j x)$     
in this expression.  We then find that  
\beqa
\p{4}{4,5}&= &-A_{4}^{(4,5)}(x) \Theta(1-4 x) + A_{3}^{(4,5)}(x) 
\Theta(1-3 x)\nonumber \\[4pt]
&& - A_{2}^{(4,5)}(x) \Theta(1-2 x)+ 
A_{1}^{(4,5)}(x) \Theta(1-x),
\label{m4n5p4}
\eeqa
where each $A_{j}^{(4,5)}(x)$ is a polynomial of 
order $mn-2 = 18$, given by 
\beqa
A_{4}^{(4,5)}(x)&=&3420 \, x (1 - 4 x)^{14} (1 + 5 x - 20 x^2 + 4 x^3),
\nonumber\\[4pt]
A_{3}^{(4,5)}(x)&=& 3420 \, x (1-3 x)^{9} \big(3 - 72 x + 552 x^2 + 360 x^3 - 19846 x^4 \nonumber \\
&& \hspace*{2 em}  + 145224 x^5 - 430948 x^6 + 580728 x^7 - 188941 x^8 \big),\nonumber \\[4pt]
A_{2}^{(4,5)}(x)&=&3420 \, x (1-2 x)^{8} \big(3 - 105 x + 1452 x^2 - 8340 x^3 + 8632 x^4 + 174904 x^5 \nonumber \\
&& \hspace*{2 em}  - 1372976 x^6 + 5366608 x^7 - 11247836 x^8 + 10332628 x^9\big),
\nonumber \\[4pt]
A_{1}^{(4,5)}(x)&=& 3420 \, x (1-x)^{11}
\big(1 - 40 x + 661 x^2 -5256 x^3
+21231 x^4 \nonumber \\
&& \hspace*{2 em}   
- 41520 x^5
+31111 x^6\big).
\label{m4n5Bi}
\eeqa
As in the case $m = n$,  
the PDFs $\p{k}{4,5}$ is written in terms of 
$A_{j}^{(4,5)}(x)$ ($j,k=1,2,3,4$) where the constants multiplying these coefficient functions are appropriate binomial coefficients with alternating signs. We get 
\beqa
\p{1}{4,5}&= &A_{4}^{(4,5)}(x) \Theta(1-4 x),\\[6pt]
\p{2}{4,5}&= &- 3 A_{4}^{(4,5)}(x) \Theta(1-4 x) + A_{3}^{(4,5)}(x) \Theta(1-3 x),\\[6pt]
\nonumber \p{3}{4,5}&=& 3 A_{4}^{(4,5)}(x) \Theta(1-4 x) - 2 A_{3}^{(4,5)}(x) 
\Theta(1-3 x)\\[4pt]
&& + A_{2}^{(4,5)}(x) \Theta(1-2 x)
\label{eqn:probsM4N5}
\eeqa
The manifest  agreement between the plots of the calculated PDFs 
and the numerical histograms validates these expressions. We have also verified that the analytical expression of the sum of the 
$q^{\rm th}$ moments of the eigenvalues 
matches the known expression~\cite{bianchi} 
for $\aver{\mathrm{Tr} (\rhoa^{(4)})^{q}}$ in this case.

\end{document}